\documentclass[doublecol]{epl2}

\usepackage{graphicx} 
\usepackage{dcolumn}
\usepackage{bm}
\usepackage{amssymb}
\usepackage{amsmath}
\usepackage{wasysym}
\usepackage{color}
\usepackage{hyperref}
\usepackage{dcolumn}
\usepackage{float}
\usepackage{flushend}
\hypersetup{
	colorlinks,%
	citecolor=blue,%
	filecolor=blue,%
	linkcolor=blue,%
	urlcolor=blue}

\title{Energy transfers and magnetic energy growth in small-scale dynamo}

\author{Rohit Kumar \inst{1} \thanks {E-mail: \email{rohitkr@iitk.ac.in}}, Mahendra K. Verma \inst{1} \thanks {E-mail: \email{mkv@iitk.ac.in}} 
\and Ravi Samtaney \inst{2} \thanks {E-mail: \email{ravi.samtaney@kaust.edu.sa}}}
\shortauthor{Rohit Kumar, Mahendra K. Verma, and Ravi Samtaney}

\institute{                    
  \inst{1} Department of Physics, Indian Institute of Technology - Kanpur 208016, India\\
  \inst{2} Mechanical Engineering, Division of Physical Sciences and Engineering, King Abdullah University of Science and Technology Thuwal  23955-6900, Kingdom of Saudi Arabia
}


\abstract{
In this letter we investigate the dynamics of magnetic energy growth in small-scale dynamo by studying energy transfers, mainly energy fluxes and shell-to-shell energy transfers.  We perform dynamo simulations for magnetic Prandtl number $\mathrm{Pm}=20$ on $1024^3$ grid using pseudospectral method.  We demonstrate that the magnetic energy growth is caused by nonlocal energy transfers  from the large-scale or forcing-scale velocity field  to small-scale magnetic field.  The peak of these energy transfers move towards lower wavenumbers as dynamo evolves, which is the reason why  the integral scale of the magnetic field increases with time. The energy transfers $U2U$ (velocity to velocity) and $B2B$ (magnetic to magnetic) are forward and local. }

\begin{document}

\maketitle

Spontaneous generation of magnetic fields in magnetohydrodynamics (MHD), particularly in stars, planets, and galaxies,  is known as ``dynamo effect"~\cite{Moffat:book}.   A small seed magnetic field is amplified by self-induced currents.  It has been argued that the swirling and twisting of the magnetic field lines lead to this growth~\cite{Vainshtein:SPU1972,Childress:book}.  In this letter we attempt to quantify the energy transfers during the dynamo process.

The magnetic energy growth is called ``small-scale dynamo'' (SSD) or ``large-scale dynamo" (LSD) depending on the scale at which magnetic field grows maximally.  Following Cattaneo {\it et al.}~\cite{Cattaneo:JFM2002}, in LSD (SSD),  the magnetic field is generated at characteristic lengths larger (smaller) than those of the velocity field.  The nature of dynamo generally depends on the magnetic Prandtl number $\mathrm{Pm}$, which is the ratio of the kinematic viscosity ($\nu$) and the magnetic diffusivity ($\eta$) of the fluid.  Typically, SSD is observed for larger $\mathrm{Pm}$~\cite{Schekochihin:APJ2004}, while LSD is for smaller $\mathrm{Pm}$~\cite{Monchaux:PRL2007}.  Yet, there are many exceptions. For example,  both SSD and LSD coexist in solar dynamo for which $\mathrm{Pm} \ll 1$~\cite{Cattaneo:JFM2002}.

Scale-dependent energy transfers in MHD have been computed using numerical and  analytical tools.  Primarily this is done by computing various energy fluxes and shell-to-shell energy transfers of MHD.  Several simulations~\cite{Dar:PD2001,Debliquy:PP2005,Carati:JT2006} employed logarithmic-binned shells, while Alexakis {\it et al.}~\cite{Alexakis:PRE2005} used linearly-binned shells.   For unit Prandtl number, these diagnostics demonstrate that under steady state, the large-scale magnetic field is fed by the large-scale velocity field.  The magnetic energy thus enhanced at large scales cascades forward to small scales.  The magnetic energy at large scales is maintained by this mechanism.  

Moll {\it et al}.~\cite{Moll:APJ2011} employed Alexakis {\it et al.}'s method~\cite{Alexakis:PRE2005} to compute energy transfers for small-scale dynamo with high $\mathrm{Pm}$.  They showed that during dynamo action, kinetic energy at large scales is transferred to the magnetic energy at smaller scales.   In our  letter we compute energy transfers using Dar {\it et al.}'s method~\cite{Dar:PD2001} for a small-scale dynamo on a high-resolution grid.  We observe that the magnetic energy growth is due to nonlocal energy transfers from large-scale velocity fields to small-scale magnetic fields.  The peak of these transfers shifts towards smaller wavenumbers with time, which leads to a growth of magnetic energy at relatively large length scales, a phenomena observed in many numerical simulations.   We also observe a forward and local magnetic-to-magnetic energy cascade, thus ruling out a proposed mechanism for the growth of the large-scale magnetic field by inverse cascade.

There are a significant number of laboratory experiments~\cite{Gailitis:PRL2000,Stieglitz:PF2001,Monchaux:PRL2007}, numerical simulations \cite{Chou:ApJ2001,Ponty:PRL2005,Mininni:PRE2005b,Yousef:PRL2008}, and shell model computations \cite{Plunian:PR2013,Lessines:PRE2009}  that address the dynamo process.   It is important to contrast the energy transfer mechanisms for SSD and LSD.  Our analysis show that in SSD, the growth of magnetic energy takes place due to a nonlocal energy transfer from large scale velocity fields to small scale magnetic fields. On the contrary, in the case of LSD, the energy transfer from the velocity field to the magnetic field is local and predominantly at large scales.


We employ pseudospectral method to solve the MHD equations, which are the governing equations for dynamo.  The non-dimensional incompressible MHD equations are~\cite{Moffat:book}
\begin{eqnarray}
\partial_{t}\mathbf{u}+ (\mathbf{u} \cdot \nabla) \mathbf{u} & = & -\nabla \left(\frac{p}{\rho}\right)+ \frac{\mathbf{J} \times \mathbf{b}}{\rho}
+ \nu \nabla^{2}\mathbf{u} + \mathbf{F}, \label{eq:MHD_vel}\\
\partial_{t}\mathbf{b}+ (\mathbf{u} \cdot \nabla) \mathbf{b} & = & (\mathbf{b} \cdot \nabla) \mathbf{u}+ \eta \nabla^{2}\mathbf{b}, \label{eq:MHD_mag}\\
\nabla \cdot \mathbf{u} & = & 0, \\ \label{eq:div_u_0}
\nabla \cdot \mathbf{b} & = & 0,  \label{eq:div_b_0}
\end{eqnarray}
where $\mathbf{u}$ and $\mathbf{b}$ are the velocity and magnetic field respectively, $\rho$ is the constant fluid density, $p$ is the thermal pressure, $\mathbf{J} = \nabla \times \mathbf b$ is the current, and $\mathbf{F}$ 
is the external force field. We perform our simulation using a pseudo-spectral code Tarang \cite{verma:Pramana2013}. The simulations have been carried out in a three-dimensional box of  size $(2\pi)^3$ with periodic boundary condition in all the three directions. The grid size for all the simulations is $1024^3$.  We employ Runge-Kutta fourth order (RK4) scheme for time stepping, and 2/3 rule for dealiasing.

In the present letter we focus on MHD simulations for high Prandtl number, for which we choose $\nu = 0.01$, $\eta =0.0005$, i.e., $\mathrm{Pm} =20$.   We apply random nonhelical (zero kinetic and magnetic helicities) forcing to the velocity field in a narrow wavenumber band $k=[2,4]$ such that the energy supply rate is maintained at a constant value.  Following the same approach as earlier work by Ponty {\it et al.}~\cite{Ponty:PRL2005}, we first perform pure fluid simulation with $\nu = 0.01$ until a steady state is reached.  For this state, the total kinetic energy is approximately 13, and the Reynolds number $\mathrm{Re}$ is approximately $666$.  The kinetic and magnetic Reynolds numbers are defined as $\mathrm{Re} = u_\mathrm{rms}L_u / \nu$ and $\mathrm{Rm} = u_\mathrm{rms}L_u / \eta$ respectively, where $u_\mathrm{rms}$ is the rms speed, and $L_u$ ($= 2\pi \int k^{-1} E_u(k) dk / \int E_u(k) dk$) is the velocity integral length scale of the system~\cite{Bhat:MNRAS2013,Ponty:PRL2005}. After this, we start our MHD simulations with two sets of initial conditions.  In the first set, the total magnetic energy of $10^{-4}$ unit is distributed uniformly over a narrow band (termed in shorthand as NB) of wavenumbers ($k=[2, 4]$), while in the second set, the initial magnetic energy of $10^{-4}$ unit is distributed uniformly over a broad band (BB; $E_b(k) \sim k^2$) of wavenumbers ($k=[2,384]$). We have carried out simulations till $t_\mathrm{final} \approx 20$ non-dimensional time units.  During the final stages,  $L_u \approx 1.5$ and $u_\mathrm{rms} \approx 1$, hence one integral-scale turnover time is approximately 1.5 time units.  Consequently, our simulation has been carried out till around 13 eddy turnover time, which is much smaller than the magnetic diffusive time $L^2/\eta \approx 8 \times 10^4$ (with $L = 2\pi$).   Note however that   the magnetic energy tends to reach close to its saturated value in 30 to 50 time units~\cite{Schekochihin:APJ2004,Cattaneo:JFM2002,Chou:ApJ2001}.   Hence our simulation starts from kinematic regime, and reaches somewhat near the saturation stage.  We study energy transfers in the dynamo starting from kinematic regime to the near saturation regime.

\begin{table}[htbp]
\caption{During the final stages of the simulation (near $t_\mathrm{final} = 20$ non-dimensional time unit), the rms speed $u_\mathrm{rms}$, the rms value of the magnetic field $b_\mathrm{rms}$, velocity integral length scale $L_u$,  magnetic integral length scale $L_b$, Reynolds number $\mathrm{Re}$, and magnetic Reynolds number $\mathrm{Rm}$.} 
\centering
\begin{tabular}{c c c c c c c}
\hline 
\hline  
Run & $u_\mathrm{rms}$ & $b_\mathrm{rms}$ & $L_u$  & $L_b$  & $\mathrm{Re}$ & $\mathrm{Rm}$ \\ [1ex]
\hline 
Pure fluid & $5.09$ & $-$ & $1.31$ & $-$ & $666$ & $-$\\
 
SSD (NB) & $0.86$ & $0.80$ & $1.67$ & $0.56$ & $143$ & $2860$\\
 
SSD (BB) & $0.92$ & $0.72$ & $1.71$ & $0.41$ & $157$ & $3140$\\
 
LSD (NB) & $1.39$ & $0.02$ & $1.29$ & $0.76$ & $896$ & $179$\\ [1ex]
\hline
\hline
\end{tabular}
\label{table:table_1} 
\end{table}

We compute energy fluxes and shell-to-shell energy transfers for the two aforementioned SSD runs at different stages of evolution, and attempt to understand the energy transfer mechanism for SSD.    To contrast with low Prandtl number simulation, we also simulate the MHD equation for $\nu = 0.002$, $\eta =0.01$ ($\mathrm{Pm} =0.2$) with the same forcing scheme and initial condition as NB SSD.

\begin{figure}[htbp]
\centering
\includegraphics[scale=0.40]{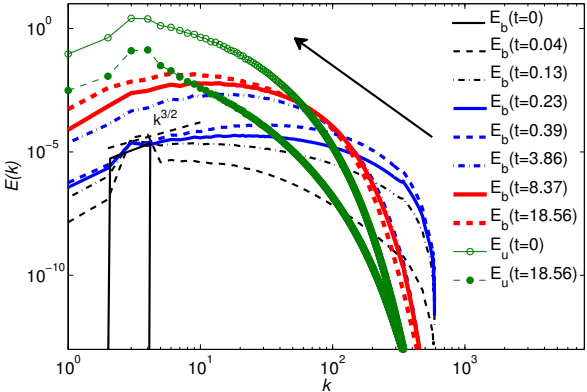} 
\caption{(Colour on-line) Evolution of kinetic ($E_u(k)$) and magnetic ($E_b(k)$) energy spectra for $\mathrm{Pm}=20$.  The initial seed magnetic field is applied in a narrow wavenumber band $k=[2,4]$ (NB). The magnetic field grows at scales smaller than the length scale of the velocity field. $E_b(k)$ quickly  spreads out in wavenumber space, after which the peak of $E_b(k)$ tends to shift leftwards.  The magnetic energy spectra in the intermediate phase of evolution show Kazantsev scaling ($E_b(k) \propto k^{3/2}$).}
\label{fig:Ebk_NB}
\end{figure}

\begin{figure}[htbp]
\centering
\includegraphics[scale=0.40]{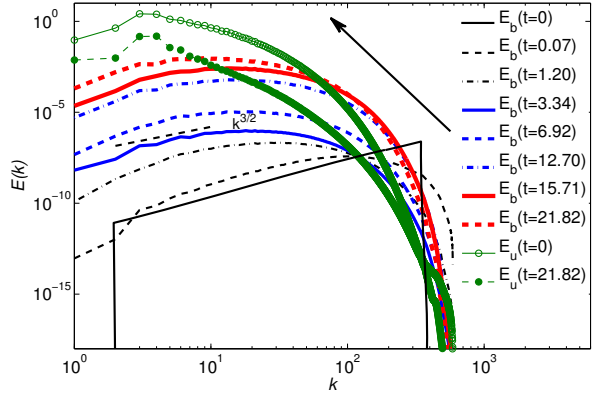}
\caption{(Colour on-line) Evolution of kinetic ($E_u(k)$) and magnetic ($E_b(k)$) energy spectra for $\mathrm{Pm}=20$ with the initial seed magnetic field applied in a broad wavenumber band $k=[2,384]$ (BB).  The evolution of $E_b(k)$ is similar to that for the NB initial condition except at very early times.}
\label{fig:Ebk_BB}
\end{figure}

Now we report the growth of the magnetic field at different stages of evolution for $\mathrm{Pm}=20$. The kinetic ($E_u(k)$) and magnetic ($E_b(k)$) energy spectra  for the initial seed field configurations, narrow-band (NB) and broad-band (BB), are illustrated in Figs.~\ref{fig:Ebk_NB} and~\ref{fig:Ebk_BB} respectively.  In the early phases, the magnetic field grows at scales smaller than  the characteristic length of the velocity field, thus indicating a presence of small-scale dynamo in these systems. As illustrated in Fig.~\ref{fig:Ebk_NB}, for the NB run, the magnetic energy gets spread out in the wavenumber space in a very short time, indicating energy transfers from the  velocity modes to all the magnetic modes.    After the aforementioned energy transfers, the subsequent magnetic energy spectra for both the initial conditions appear to be same.    Near the final stage, the magnetic energy tends to grow at smaller wavenumbers.  In the intermediate growth phase, $E_b(k) \propto k^{3/2}$, as predicted by the Kazantsev model of dynamo~\cite{Kazantsev:JETP1968}.  

\begin{figure}[htbp]
\centering
\includegraphics[scale=0.40]{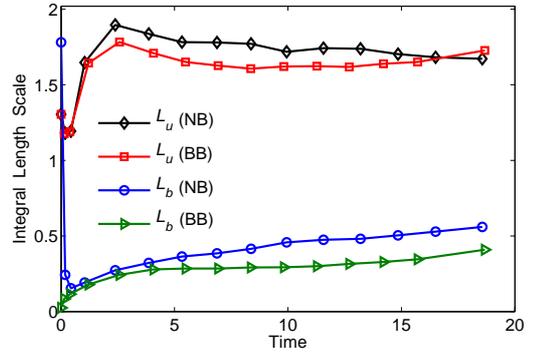} 
\caption{(Colour on-line) Plots exhibiting velocity and magnetic integral length scales ($L_u, L_b$) as a function of time for $\mathrm{Pm=20}$ with the initial conditions NB (narrow band) and BB (broad band).  $L_b$ grows with time, while $L_u$ grows first and then saturates.  In the final phase of our simulation, $L_u/L_b \approx 3$.}
\label{fig:integ_len_time}
\end{figure}

To investigate the shift of magnetic energy to small wavenumbers, we compute  the velocity integral length scale and the magnetic integral length scale ($L_b = 2\pi \int k^{-1} E_b(k) dk / \int E_b(k) dk$). The plots of integral length scales with time are shown in Figure~\ref{fig:integ_len_time}, which demonstrates that $L_b$ decreases abruptly for the NB initial condition, which is due to the aforementioned shift of magnetic energy to intermediate and small length scales.  After this brief phase, both NB and BB initial conditions exhibit growth of $L_b$,  which reflects a presence of SSD.  The growth of  $L_b$ or a corresponding decrease of equivalent wavenumber $k_b$ with time is consistent with the aforementioned shift of the magnetic energy to smaller wavenumbers.  The velocity integral scale $L_u$ however first increases then saturates. Note that $L_u/L_b \approx 3$ near $t_\mathrm{final}$, consistent with the recent results of Bhat and Subramanian~\cite{Bhat:MNRAS2013}. 

\begin{figure}[htbp]
\centering
\includegraphics[scale=0.40]{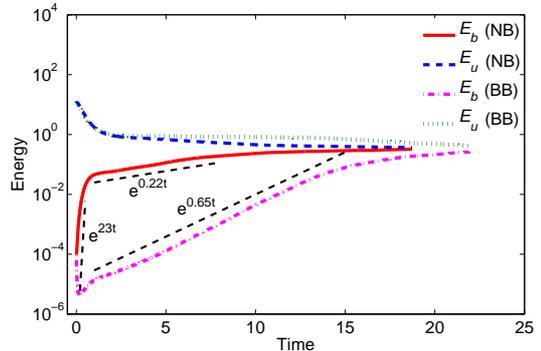} 
\caption{(Colour on-line)  Evolution of kinetic energy ($E_u$) and magnetic energy ($E_b$) for $\mathrm{Pm}=20$ when the initial magnetic energy is distributed uniformly in a narrow band (NB: $k=[2,4]$), and when it is distributed in a broader band (BB: $k=[2,384]$).  The magnetic energy grows exponentially in two phases for NB, and in one phase for BB.}
\label{fig:kin_mag}
\end{figure}

In Fig.~\ref{fig:kin_mag}, we plot the evolution of magnetic and kinetic energies as a function of time.  The magnetic energy grows for both NB and BB initial conditions.  The NB case has two phases of magnetic energy growth.  However, the BB initial condition has only one growth phase after a sharp drop of $E_b$ at very early times, which could be a transient.  The first phase of the $E_b$ growth ($E_b \sim \exp(23t)$) for the NB initial condition is due to the aforementioned rapid energy transfer to the magnetic field at intermediate and small scales in the early phases, while the second phase of $E_b$ growth corresponds to the predictions of Kazantsev model~\cite{Kazantsev:JETP1968}; these results are similar to those of Chou~\cite{Chou:ApJ2001} and Schekochihin {\it et al.}~\cite{Schekochihin:APJ2004}.  Note that in this regime, the exponents for the NB and BB initial conditions are different, and they are 0.65 and 0.22 respectively.  This is consistent with earlier numerical results of Cattaneo {\em et al.}~\cite{Cattaneo:JFM2002} and Schekochihin {\it et al.}~\cite{Schekochihin:APJ2004}, according to which the dynamo growth depends quite sensitively on the initial condition. 

The integral length of the magnetic field and the magnetic energy appear to flatten out near the final stages of our simulations (near $t_\mathrm{final} = 20$). However, the final states of our simulations are at some distance away from the final asymptotic regime because the above quantities have not yet saturated.  We had to stop our simulations before reaching the asymptotic state due to  excessive computing time required to carry out these simulations for $1024^3$ grids.

After the discussion on the total energy and energy spectra, we turn to the flux computations, which is one of the main topics of this letter.  The energy flux from the region  $X$ (of wavenumber space) of $\alpha$ field  to the region  $Y$ of $\beta$ field is given by \cite{Dar:PD2001, Verma:PR2004,Debliquy:PP2005} 
\begin{eqnarray}
\Pi^{\alpha,X}_{\beta,Y} = \displaystyle\sum_{\mathbf{k} \in Y} \displaystyle\sum_{\mathbf{p} \in X} S^{\beta \alpha} (\mathbf{k} | \mathbf{p} |\mathbf{q}), \label{eq:flux}
\end{eqnarray}
where the ``mode-to-mode energy transfer rate" $S^{\beta \alpha} (\mathbf{k}| \mathbf{p}| \mathbf{q})$ represents energy transfer from mode $\mathbf{p}$ of $\alpha$ field to mode $\mathbf{k}$ of $\beta$ field with the mode $\mathbf{q}$ acting as a mediator.  Note that the triad ($ {\mathbf {k,p,q}}$) satisfies a condition  $\mathbf{k} + \mathbf{p} + \mathbf{q} = 0$.     Here we provide an expression for $S^{bu} (\mathbf{k} | \mathbf{p} | \mathbf{q})$, the energy transfer from $\mathbf u(\mathbf p)$ to  $\mathbf b(\mathbf k)$, as an illustration:
\begin{eqnarray}
S^{bu} (\mathbf{k} | \mathbf{p} | \mathbf{q}) = \Im ([\mathbf{k} \cdot \mathbf{b} (\mathbf{q})][\mathbf{b} (\mathbf{k}) \cdot \mathbf{u} (\mathbf{p})]), \label{eq:mode-to-mode}
\end{eqnarray}
where $\Im$ denotes the imaginary part of the argument. 
Fluid turbulence involves only one energy flux $\Pi^{u<}_{u>}(k_0)$, which is defined as the energy transfer from the modes residing inside a wavenumber sphere of radius $k_0$ to the modes residing outside the sphere.   Here $<$ and $>$ represent the modes residing inside and outside respectively.   MHD turbulence however has six energy fluxes: $\Pi^{u<}_{u>}(k_0)$, $\Pi^{u<}_{b>}(k_0)$, $\Pi^{b<}_{b>}(k_0)$, $\Pi^{b<}_{u>}(k_0)$, $\Pi^{u<}_{b<}(k_0)$, and $\Pi^{u>}_{b>}(k_0)$. Dar {\it et al.}~\cite{Dar:PD2001} and Verma \cite{Verma:PR2004} have constructed formulas to compute these fluxes.  Here we quote only one of them.  The energy flux from inside of the $u$-sphere of radius $k_0$ to outside of the $b$-sphere of the same radius is  
\begin{eqnarray}
\Pi^{u<}_{b>}(k_0) = \displaystyle\sum_{|\mathbf{k}| > k_0} \displaystyle\sum_{|\mathbf{p}| < k_0} S^{b u} (\mathbf{k} | \mathbf{p} | \mathbf{q}). \label{eq:flux_ub}
\end{eqnarray}

\begin{figure*}[htbp]
\centering
\includegraphics[scale=0.30]{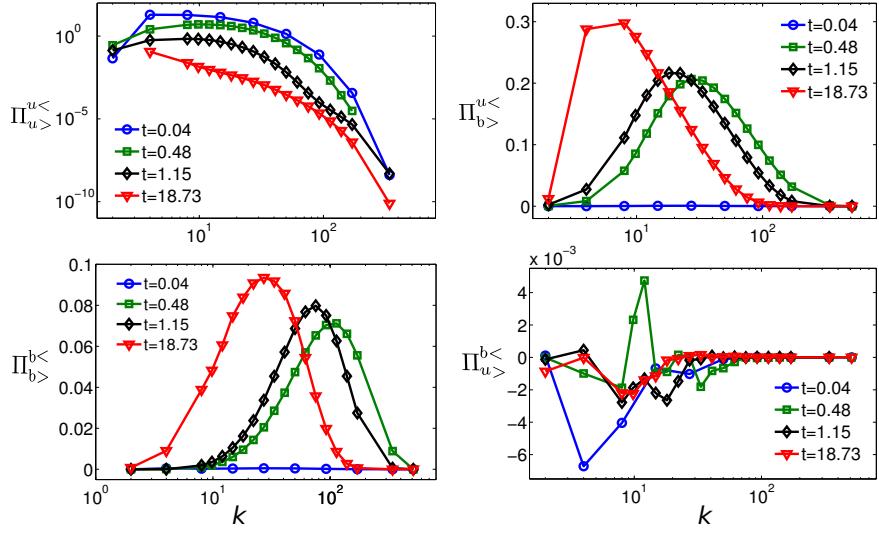}
\caption{(Colour on-line) Plots of energy fluxes $\Pi^{u<}_{u>}$, $\Pi^{u<}_{b>}$, $\Pi^{b<}_{b>}$, and $\Pi^{b<}_{u>}$ vs. $k$.  The first three fluxes are forward, while $\Pi^{b<}_{u>}$ comes with both positive and negative signs.  Note that $\Pi^{u<}_{b>}$ (from velocity to magnetic) dominates all other fluxes.  Also, $\Pi^{u<}_{b>}$, $\Pi^{b<}_{b>}$ at $t=0.04$ are very small, but nonzero.}
\label{fig:ssd_flux}
\end{figure*}

In Figure~\ref{fig:ssd_flux} we exhibit various energy fluxes computed using NB simulation data. The energy flux from inner $u$-sphere to outer $u$-sphere ($\Pi^{u<}_{u>}$), inner $b$-sphere to outer $b$-sphere ($\Pi^{b<}_{b>}$), and inner $u$-sphere to outer $b$-sphere ($\Pi^{u<}_{b>}$) are all positive.   Positive value of $\Pi^{b<}_{b>}$ implies that the growth of magnetic energy at larger scales (leftward shift of the $E_b$ peak) is not due to any inverse cascade of magnetic energy, as conjectured by some of the earlier work~\cite{Pouquet:JFM1976,Plunian:PR2013}.   The leftward shift of the peak of the $E_b(k)$ however is due to the shift of dominant flux $\Pi^{u<}_{b>}$ to smaller wavenumbers, as exhibited in Figure~\ref{fig:ssd_flux}.   The energy flux $\Pi^{b<}_{u>}$, which  has both signs, is much smaller compared to the aforementioned energy fluxes.  We also remark that the energy fluxes for BB initial condition have similar behaviour. Another important point to note is that all the fluxes are still evolving at $t \approx 20$, and they have not reached the final steady state, yet they provided valuable information about the energy transfers.

The energy fluxes described above provide information on the cumulative energy transfers.  To obtain a more refined view of the energy transfers responsible for SSD, we compute the shell-to-shell energy transfer rates.   In MHD, there are three kinds of shell-to-shell energy transfer rates \cite{Dar:PD2001,Verma:PR2004}:  from velocity to velocity field ($U2U$), from magnetic to magnetic ($B2B$), and from velocity to magnetic ($U2B$). The shell-to-shell energy transfer from the $m$th shell of $\alpha$ field  to the $n$th shell of $\beta$ field is defined as~\cite{Debliquy:PP2005, Dar:PD2001, Verma:PR2004} 
\begin{eqnarray}
T^{\beta,\alpha}_{n,m} = \displaystyle\sum_{\mathbf{k} \in n} \displaystyle\sum_{\mathbf{p} \in m} S^{\beta \alpha} (\mathbf{k} | \mathbf{p} | \mathbf{q}). \label{eq:shell-to-shell}
\end{eqnarray}
For example, the shell-to-shell energy transfer rate from the $m$th shell of $u$ field to the $n$th shell of $b$ field is  
\begin{eqnarray}
T^{b,u}_{n,m} = \displaystyle\sum_{\mathbf{k} \in n} \displaystyle\sum_{\mathbf{p} \in m} S^{b u} (\mathbf{k} | \mathbf{p} | \mathbf{q}) \label{eq:shell-to-shell_ub}
\end{eqnarray}

\begin{figure}[htbp]
\centering
\includegraphics[scale=0.45]{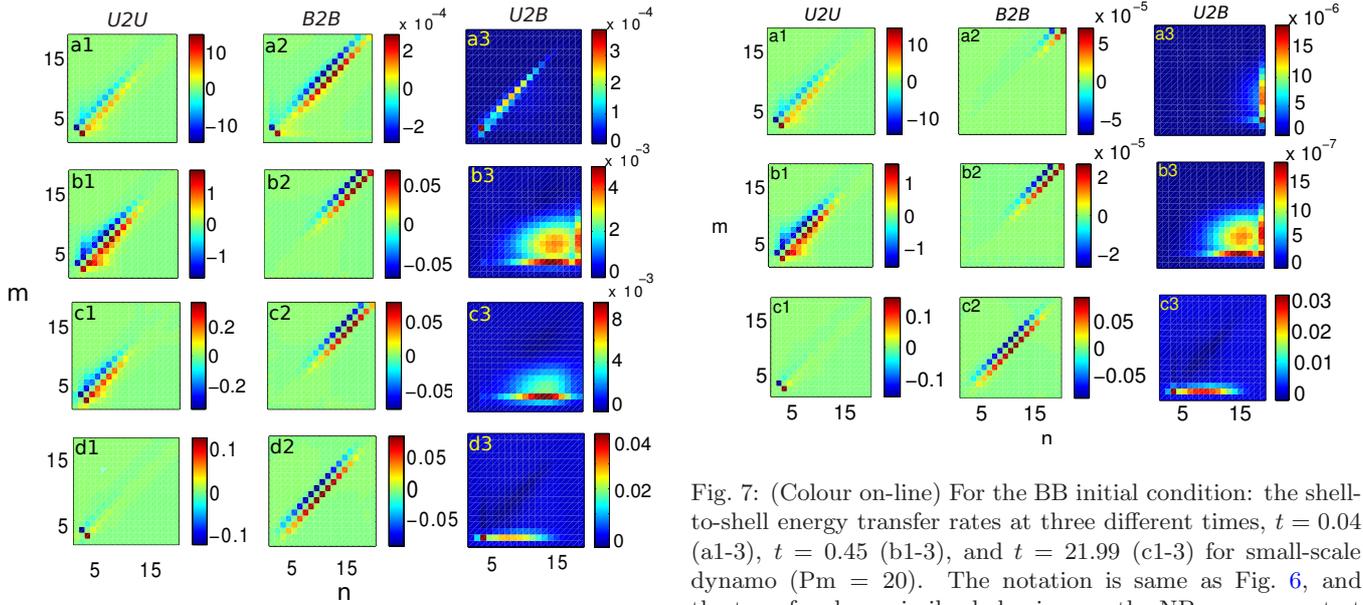} 
\caption{(Colour on-line) Shell-to-shell energy transfer rates for small-scale dynamo ($\mathrm{Pm} =20$) with NB initial condition:  $U2U$ (velocity to velocity), $B2B$ (magnetic to magnetic), and $U2B$ (velocity to magnetic) at four different times, $t=0.04$ (a1-3),  $t=0.48$ (b1-3),  $t=1.15$ (c1-3),  and $t=18.73$ (d1-3).  Here the horizontal axes represent the receiver shells, while the vertical axes represent the giver shells.   The energy transfers  $U2U$ and $B2B$ are local and forward, but $U2B$ is forward and nonlocal (except at $t=0.04$).  }
\label{fig:shell_uu_bb_ub}
\end{figure}

\begin{figure}[htbp]
\centering
\includegraphics[scale=0.45]{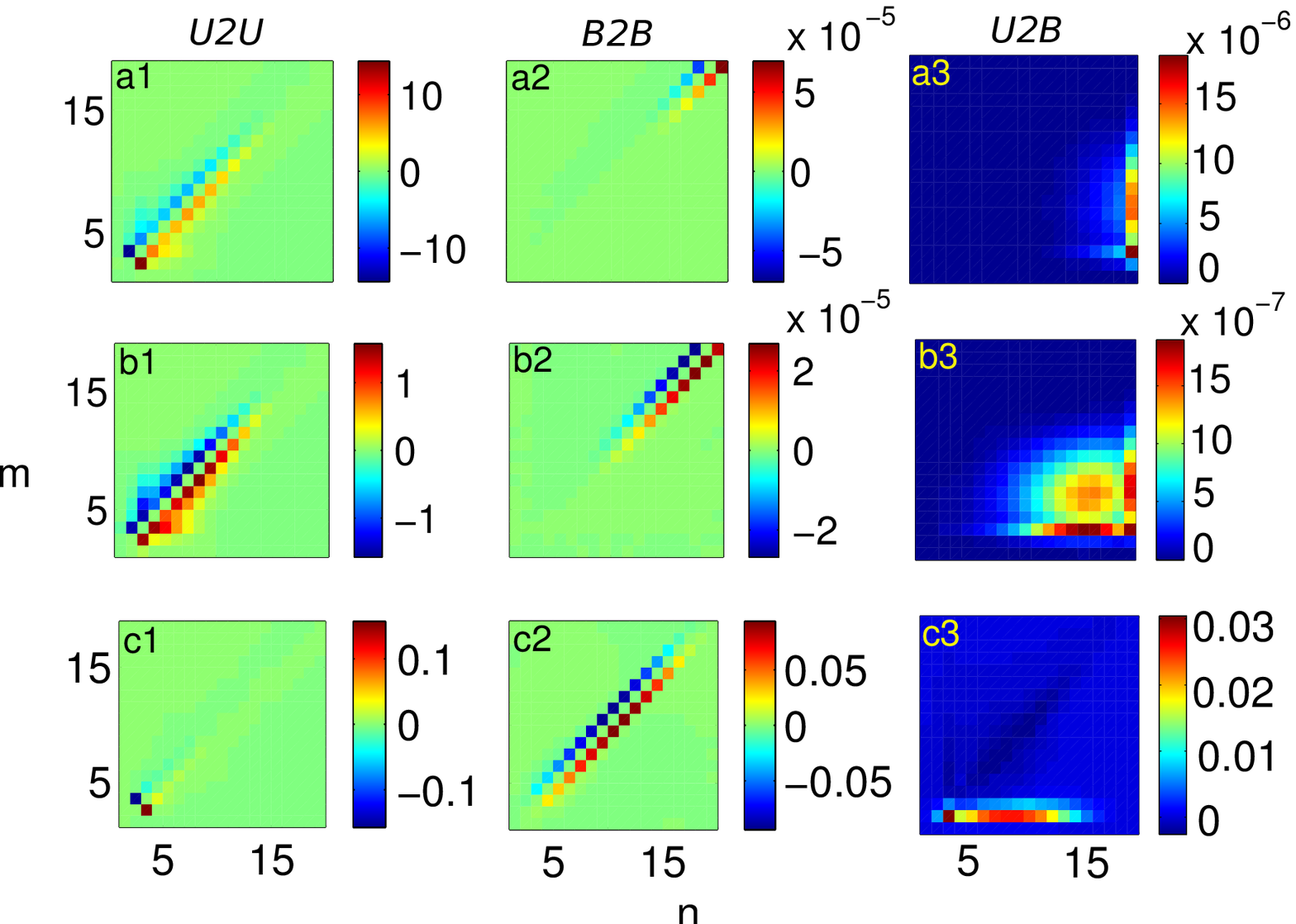}
\caption{(Colour on-line) For the BB initial condition: the shell-to-shell energy transfer rates at three different times, $t=0.04$ (a1-3),  $t=0.45$ (b1-3), and $t=21.99$ (c1-3) for small-scale dynamo ($\mathrm{Pm} =20$).  The notation is same as Fig.~\ref{fig:shell_uu_bb_ub}, and the transfers have similar behaviour as the NB case, except at $t=0.04$ when the $U2B$ shell-to-shell transfer is nonlocal (from all $u$ shells to the last $b$ shell).} 
\label{fig:shell_uu_bb_ub_flat}
\end{figure}

The shell-to-shell energy transfer rates for the NB initial condition  is shown in Figure~\ref{fig:shell_uu_bb_ub}, while those for  BB case is shown in Figure~\ref{fig:shell_uu_bb_ub_flat}.  In both the cases, the radii of the wavenumber shells are: 
$2$, $4$, $8$, $9.8$, $12$, $14.8$, $18.1$, $22.2$, $27.2$, $33.4$, $40.9$, $50.2$, $61.5$, $75.4$, $92.5$, $113.4$, $139$, $170.5$, $341$, and $512$.   The $U2U$ and $B2B$ energy transfers are  forward, i.e., the energy is transferred from smaller wavenumbers to larger wavenumbers.  Also, the energy transfers are local, i.e., dominantly among the neighbouring wavenumber shells.  The $U2U$ energy transfers are large  in the initial stage, but their  magnitudes decrease with time.   Also, at later phases $E_u$ is concentrated at smaller wavenumbers  (non-Kolmogorov) since  $\mathrm{Re}$ becomes relatively low in this regime. Consequently, near $t=t_\mathrm{final}$, the $U2U$ shell-to-shell transfer is significant only for small $n$, as exhibited in Figs.~\ref{fig:shell_uu_bb_ub}(d1) and \ref{fig:shell_uu_bb_ub_flat}(c1).  

One of the most interesting features of the shell-to-shell transfers is that the $U2B$ energy transfers are forward as well as nonlocal, except in the very early stages (e.g., in Fig.~\ref{fig:shell_uu_bb_ub}(a3) at $t=0.04$).   Note that the  $U2B$  transfers involve interactions among velocity and magnetic modes~\cite{Dar:PD2001,Debliquy:PP2005,Carati:JT2006,Verma:PR2004,Verma:PP2005}.  Therefore, at later phase, the $U2B$ transfer is  nonlocal because the velocity modes dominate at small wavenumbers, while the magnetic modes at large wavenumbers; the small-$k$ $u$ modes interact with large-$k$ $B$ modes.  We observe that near $t\approx 0$, the $U2B$  transfers are local for the NB case since $E_b(k)$ is not spread out as much.  On the contrary, for the BB  case, the last magnetic shell receives energy from all the velocity shells since $E_b(k)$ peaks at the last shell.    Our computations also reveal that the peak of nonlocal $U2B$ transfer shifts towards lower wavenumbers as dynamo evolves, which is the reason for the shift of magnetic energy to smaller wavenumbers with time.  We must however point out that the nonlocality in $U2B$ energy transfer could get significant contributions from a possible correlations between the $U$ and $B$ fields induced by forcing~\cite{Carati:JT2006}.

\begin{figure}[htbp]
\centering
\includegraphics[scale=0.56]{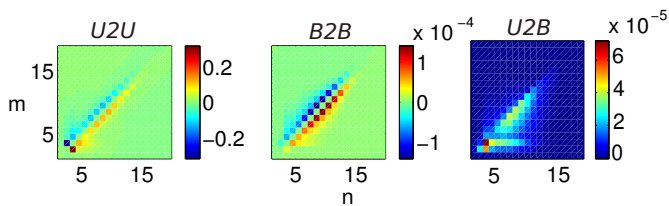} 
\caption{(Colour on-line) Shell-to-shell energy transfer rates for large-scale dynamo ($\mathrm{Pm} =0.2$) with NB initial condition.  The plots are for $t=3.69$, which is the intermediate time during the magnetic energy growth.  The notation is same as Fig.~\ref{fig:shell_uu_bb_ub}.  The three energy transfers  $U2U$, $B2B$, and $U2B$ are local and forward.}
\label{fig:lsd_shell_uu_bb_ub}
\end{figure}

To contrast the energy transfers mechanisms between SSD and LSD, we compute shell-to-shell energy transfers for $\mathrm{Pm} =0.2$, which is a sample of low Prandtl number.   In Fig.~\ref{fig:lsd_shell_uu_bb_ub} we present $U2U, B2B$ and $U2B$ shell-to-shell transfers for this case at $t=3.69$, which is in the intermediate stage of the $E_b$ growth.  The $U2U$ and $B2B$ transfers are local as expected. The $U2B$ transfers are  also predominantly local, but with a weak nonlocal component, which could be due to field correlations induced by the external forcing~\cite{Carati:JT2006}. The predominantly local nature of energy transfers is due to the fact that both $E_u(k)$ and $E_b(k)$ are significant at small wavenumbers, unlike SSD case in which $E_b(k)$ is significant only at large wavenumbers.  A detailed comparison between the energy transfers in SSD and LSD will be presented in a future article.

The above results on energy transfers demonstrate that the energy fluxes and shell-to-shell energy transfer rates provide very valuable information on the growth mechanism of the magnetic field.   We show that for high magnetic Prandtl numbers, the kinetic energy tends to be concentrated in the smaller and intermediate wavenumber regions.  However, the magnetic field is spread out in the wavenumber space with large wavenumber modes containing significant magnetic energy.  As a result, a nonlocal energy transfer takes place from small wavenumber velocity modes to large wavenumber magnetic modes ($U2B$).   This process is responsible for the growth of the magnetic field at large wavenumbers or small length scales. These results are consistent with the Moll {\it et al}.'s findings~\cite{Moll:APJ2011}, yet our simulations provide further insights into locality as well as magnitudes of energy transfers.   Our findings are contrary to the earlier conjectures that the shift of the magnetic energy peak to smaller wavenumbers  (for nonhelical dynamos) is due to an inverse cascade of magnetic energy from small length scales to large length scales~\cite{Pouquet:JFM1976, Plunian:PR2013}.  We also point out that energy transfer processes for  $\mathrm{Pm}$ near unity~\cite{Dar:PD2001,Debliquy:PP2005,Alexakis:PRE2005,Carati:JT2006,Moll:APJ2011} have significant differences with those for $\mathrm{Pm} = 20$ shown here.  We also show that the above energy transfers at final stages are qualitatively similar for two different seed fields (narrow band at low wavenumbers, and uniform broad band), but the growth rate of the magnetic energy depends on the initial condition.

In summary, the energy transfer studies of small-scale dynamo provide valuable insights into the dynamics of magnetic energy growth. A generalization of this analysis to large-scale dynamo would be very valuable for understanding dynamo mechanism.


\begin{acknowledgments}
 We thank K. Subramanian  for valuable discussions and comments, and Rakesh Yadav for performing initial set of numerical simulations. We are grateful to the anonymous referees for comments and suggestions that helped us improve the manuscript. Our numerical simulations were performed on the IBM Blue Gene P ``Shaheen'' at KAUST supercomputing laboratory, Saudi Arabia. This work was supported through the Swarnajayanti fellowship to MKV from Department of Science and Technology, India. RS was supported through baseline funding at KAUST. 
\end{acknowledgments}



\end{document}